Title: Monitoring through many eyes: Integrating disparate datasets to improve monitoring of the Great Barrier Reef

Authors: Erin E Peterson[a,b,c,l], Edgar Santos-Fernández[b,c], Carla Chen[d], Sam Clifford[b,c], Julie Vercelloni[b,e,l], Alan Pearse[a,b,l], Ross Brown[f,l], Bryce Christensen[a,l], Allan James[a,l], Ken Anthony[d], Jennifer Loder[g,h], Manuel González-Rivero[d,i], Chris Roelfsema[j], M. Julian Caley[b,c], Camille Mellin[d,k], Tomasz Bednarz[b,l], and Kerrie Mengersen[b,c,l]

Affiliations:

[a] Institute for Future Environments, Queensland University of Technology, Brisbane, Australia

[b] Australian Research Council Centre of Excellence for Mathematical and Statistical Frontiers (ACEMS)

[c] School of Mathematical Sciences, Queensland University of Technology, Brisbane, Australia

[d] Australian Institute of Marine Science, Townsville, Australia

[e] Australian Research Council Centre of Excellence for Coral Reef Studies, School of Biological Sciences, The University of Queensland, St Lucia, Australia

[f] School of Electrical Engineering and Computer Science, Queensland University of Technology, Brisbane, Australia

[g] Reef Check Australia, Brisbane, Australia

[h] Reef Citizen Science Alliance, Brisbane, Australia

[i] Global Change Institute, The University of Queensland, Brisbane, Australia

[j] Remote Sensing Research Centre, School of Earth and Environmental Science, The University of Queensland, Brisbane, Australia

[k] The Environment Institute and School of Biological Sciences, University of Adelaide

[l] The Cooperative Research Centre for Spatial Information, Docklands, Australia

**Abstract:**

Numerous organisations collect data in the Great Barrier Reef (GBR), but they are rarely analysed together due to different program objectives, methods, and data quality. We developed a weighted spatio-temporal Bayesian model and used it to integrate image-based hard-coral data collected by professional and citizen scientists, who captured and/or classified underwater images. We used the model to predict coral cover across the GBR with estimates of uncertainty; thus filling gaps in space and time where no data exist. Additional data increased the model's predictive ability by 43%, but did not affect model inferences about pressures (e.g. bleaching and cyclone damage). Thus, effective integration of professional and high-volume citizen data could enhance the capacity and cost-efficiency of monitoring programs. This general approach is equally viable for other variables collected in the marine environment or other ecosystems; opening up new opportunities to integrate data and provide pathways for community engagement/stewardship.

1. Introduction

Australia's Great Barrier Reef (GBR) is the largest coral-reef ecosystem on earth, as well as one of the most biodiverse. It includes a wide array of habitats, which support over 1500 species of fish, around 400 species of hard coral, and close to 8,000 other species of invertebrates and vertebrates including marine mammals, sea turtles and birds (GBRMPA 2009). The GBR also provides a suite of ecosystem services, such as coastline protection from wave exposure, food for people, recreational and cultural heritage benefits, and significant economic benefits to the Australian economy including tourism and fishing (Stoeckl et al. 2011; Deloitte Access Economics 2017). The GBR was designated a World Heritage Area in 1981 and the Australian Federal and Queensland State governments are committed to managing this unique natural asset using best practices. The Reef 2050 Long-Term Sustainability Plan includes targets, objectives, and outcomes for GBR management (Commonwealth of Australia 2015), while the Reef Integrated Monitoring and Reporting (RIMReP) program is tasked with integrating monitoring data and model outputs from multiple organisations to measure progress towards those outcomes (Commonwealth of Australia 2018). The goal of this study is to support those integration tasks by demonstrating how existing data from multiple organisations can be combined within a statistical model, which accounts for differences in survey method and quality, to provide spatially and temporally explicit information used to inform science and improve management.

Monitoring the GBR World Heritage Area is challenging because it is approximately 348,000 km$^2$ in size and contains 2900 coral reefs with a surface area of about 25,000 km$^2$ (Hedge et al. 2013). This makes comprehensive monitoring of reef health using traditional survey methods economically and logistically difficult. There is, however, a wealth of data available from a large number of designed and opportunistic studies within the GBR. A recent review identified more than 90 publicly and privately funded monitoring programs in operation, run by research institutions, government agencies, reef-based industries, citizen science groups, and traditional owners collecting various types of data for different purposes (GBRMPA and Queensland Government 2018). These data represent a significant investment of effort and money. The challenge is that these data vary with respect to source, type and quality, and cover different snapshots of time and space, making it difficult to combine them within a single analysis.

One of the most common indicators of reef health is hard coral cover (i.e. coral cover), which represents the percent or proportion of the benthic zone covered in hard corals, without accounting for overlap in the three-dimensional coral structure. Coral cover is often used in coral-reef monitoring programs because hard corals build critical reef habitat, are influenced by a range of disturbances (e.g. destructive fishing practices, land- or ship-based impacts, tourism and recreation, storm damage, and climate change), and consistent monitoring methodologies are available (Hill and Wilkinson 2004). However, these same characteristics make coral cover monitoring challenging, especially at broad spatial scales. For example, coral reefs have evolved within a natural environment that includes cyclone and storm damage, disease outbreaks and Crown-of-Thorns Starfish (CoTS) predation, and so it is expected that healthy coral reefs will vary spatially and temporally (Osborne et al. 2011; Vercelloni et al. 2017a). However, if the frequency and intensity of natural disturbance regimes increase, corals become more vulnerable to stress from anthropogenic disturbances such as land-based pollution, bleaching, and ocean acidification (De'ath et al. 2012; Brodie and Waterhouse 2012; Hughes et al. 2018). Recent studies show that these pressures negatively affect the ability of the coral to recover from the cumulative impacts of these disturbances (Hughes et al. 2003; Osborne et al. 2011; Hughes et al. 2017; Vercelloni et al. 2017a; Ortiz et al. 2018). Thus, the challenge is to obtain enough data to measure long-term status and trends, as well as short-term impacts at local scales throughout the GBR in order to understand the underpinning science, as well as prioritise management actions and assess whether those actions are having the desired effect.

One of the longest-standing programs for monitoring coral cover is the Australian Institute of Marine Science (AIMS) Long-term Monitoring Program (LTMP). The LTMP has measured coral cover at 47 reefs since 1993 and an additional 56 Representative Areas Program (RAP) reefs since 2006 in order to assess status and trends in coral reef condition (Delean and De'ath 2008; Sweatman et al. 2008). These data provide a resource for estimating trends in coral cover and represent the "gold standard" in terms of data quality. Although the program was designed to be spatially representative in terms of coral communities, it was not designed to capture spatial variability in coral cover, or to generate fine-scale predictions of coral cover over time, at the GBR scale. This is also true of the other professional, coral-reef monitoring programs run by research institutions and government agencies, which are individually too constrained in space or time to meet both short- and long-term management needs (Hedge et al. 2013). At the same time, thousands of surveys are undertaken each year by community groups and private organisations, but there is currently no framework to formally integrate these data with those from professional programs.

The aim of this study was to demonstrate how coral cover data from a diverse array of public and privately funded monitoring programs can be combined to provide spatially and temporally explicit information throughout the GBR, which can be used to support data-enabled management decisions. We used a mechanistic weighting scheme to account for methodological differences in coral cover data from different programs and weighted spatio-temporal Bayesian models to predict coral cover, with estimates of uncertainty, throughout the GBR and over time. More specifically, we investigated whether the inclusion of multiple data sources increased the predictive ability of the models and/or decreased the uncertainty in predictions in areas with and without existing professional monitoring data. We also explored the influence of citizen-contributed data in the model and investigated the impact that increased participation of citizens would have on model outputs. We conclude with an in-depth discussion about the potential for an integrated approach to monitoring, evaluation, and reporting of coral cover condition using this method, including monitoring the effectiveness of management actions and point out how this approach could be applied to coral reefs and other ecosystems more broadly.

**Methods**

2.1 Hard coral cover measurements

Estimates of hard coral cover are often based on transects of individual images of the benthos, which are captured using a still or video camera oriented perpendicular to the seabed. The images are then either manually annotated (i.e. classified) by marine scientists or automatically classified using software such as CoralNet (Beijbom et al. 2015) to document percent cover of hard coral. We used coral cover data from a number of different sources including the: XL Catlin Seaview Survey (González-Rivero et al. 2014); Great Barrier Reef Long-Term Monitoring Program (LTMP) and the Reef Rescue Marine Monitoring Program (MMP), conducted by the Australian Institute of Marine Science (AIMS); and the University of Queensland Remote Sensing Research Centre surveys (UQ-RSRC; Roelfsema et al. 2018a; Roelfsema et al. 2018b). Each dataset provided multiple estimates of coral cover, but there were differences in the scale of the estimates and the estimation methods (Table 1). For additional details about how the data were collected, please see Appendix 1.

2.2. Covariate data

A reference raster with a spatial resolution of 0.005 decimal degrees (dd) was created (approximately 500m$^2$), covering the extent of the reefs in the GBR. In addition, 85529 unsampled prediction locations were generated at the centroid of each reference raster cell.

A number of physical, topographic and disturbance covariates were included in the model to account for direct and indirect sources of variation in coral cover (Table 2). The covariate rasters were resampled (i.e. the spatial resolution was altered) to match the spatial resolution of the reference raster (0.005 dd). Covariate values were then extracted for all of the observed and prediction locations for inclusion in the model. Please see Appendix 2 for a description of the relationship between coral cover and the potential covariates and details about the geo-processing operations used to create them.

Table 1. Differences in the coral-cover data sources, *s*, including the number of estimates used in the modelling (N), scale of the coral cover estimate, number of images the estimate was based on, extent of each individual image, classification method, and number of classification points per image. These factors were used to derive a weight for each coral cover estimate, by source.

| Source | N | Scale | Number of images ($w_{N_s}$) | Image extent (m$^2$) ($w_{e_{js}}$) | Classification method | Classification points ($w_{n_j}$) | Coral cover estimate weight ($w_{ms}$) |
|---|---|---|---|---|---|---|---|
| UQ-RSRC[1] | 18077 | Image | 1 | 1.00 | Manual | 24 | 10 |
| XL Catlin[2] | 42386 | Image | 1 | 1.00 | Automated | 50 | 10 |
| LTMP[3] | 16851 | 5 × 50m transects | 40 | 0.20 | Manual | 5 | 40 |
| MMP[4] | 6068 | 5 × 20m transects | 32 | 0.20 | Manual | 5 | 32 |
| Reef Check Australia | 218 | Image/citizen | 1 | 0.12 | Manual | 20 | Variable |

[1] University of Queensland Remote Sensing Research Centre, [2] XL Catlin Seaview Survey, [3] Long-term Monitoring Program, [4] Marine Monitoring Program

2.3 Citizen-contributed images

A total of 218 citizen-contributed images were sourced from Reef Check Australia (i.e. Reef Check; http://www.reefcheckaustralia.org), a citizen science organisation working to monitor Queensland reefs using a globally standardised methodology. Reef Check recruits and trains volunteer divers to conduct underwater visual-reef surveys, including collecting benthic cover data, using point-intercept transects. From 2003-2009, video was also recorded along a series of transects, with the video camera oriented down directly towards the seabed. We extracted images from video footage taken at dive sites from Magnetic Island to Osprey Reef ensuring that at least 10 seconds had elapsed between them. The geographic position of each image was estimated based on the timestamps of the sampled frames, and information about the start location and bearing of the transect. Please see Appendix 1 for additional details.

Table 2. Covariates that were included in the coral cover model. The spatial resolution is given in decimal degrees.

| Covariate | Description | Source | Spatial Resolution | Temporal Resolution |
|---|---|---|---|---|
| Cyclone exposure | Hours of exposure to damaging waves caused by tropical storms/cyclones (>4m):<br><br>0= No cyclone effects,<br>1 = Some cyclone effects | Puotinen et al. (2016) | 0.01° | 2002-2015 |
| Bleaching exposure | 0= No coral bleaching, 1 = > 1% coral bleached | Matthews et al. (2019) | 0.01° | 2002 |
| Sea surface temperature anomaly | Difference between measured sea surface temperature (SST) and monthly long-term mean SST (°C) | BOM (2014) | 0.02° | Annual means for 2002-2015 |
| Shelf position | Position of reefs on the continental shelf;<br>1= inshore/inner shelf;<br>2 = middle shelf;<br>3 = outer shelf | GBRMPA (2014) | 0.005° | Great Barrier Reef Zoning Plan 2003 |
| No Take Zone | Protected areas where no fishing is allowed. 1 = no-take, 0 = otherwise | GBRMPA (2014) | 0.005° | Great Barrier Reef Zoning Plan 2003 |

2.4 Eliciting coral cover using citizen scientists

An elicitation (i.e. classification) tool was developed to allow citizens to browse a map of the GBR and select images for classification (www.virtualreef.org.au). There were a total of 12 citizen scientists who took part in this study. This included 6 members of the project team (a subset of the co-authors) with no expertise in marine science or hard coral identification. The other 6 citizen scientists were members of Reef Check, who had undertaken multi-day classroom and field-based training in a standardised, reef-health monitoring protocol, including identifying hard corals to growth form. Data collected by Reef Check have been successfully used to detect major trends in coral cover (Done et al. 2017), which suggests that trained surveyors have the skills to identify hard corals in images.

We used the classifications contributed by citizens in the model in two ways and this affected the way the classification was performed. For the 218 Reef Check images, a unique spatially balanced random sample of 20 classification points was generated for each citizen, which they classified as either water, (hard) live coral, algae, sand, unknown, or other (Figure 1). Once all 20 points were annotated, the user submitted the classifications to a database, where the image's media identifier (ID), latitude and longitude, a classification identification (ID) number (for each point in the image), classification label, and the user ID for the citizen providing the classification were recorded. For each image and each citizen, a coral cover estimate was obtained based on the number of points labelled "coral" as a fraction of the number of points labelled as something other than "unknown". Citizens were also asked to annotate an additional 20 Catlin images, which we used as validation images. We asked a marine scientist with expertise in coral reef ecology to classify the 20 validation images and then asked each citizen to classify the *same* classification points within the images. This allowed us to assess each individual's ability to accurately identify coral in the imagery. We describe how this accuracy measure was calculated and is used in Section 2.5.1.2.

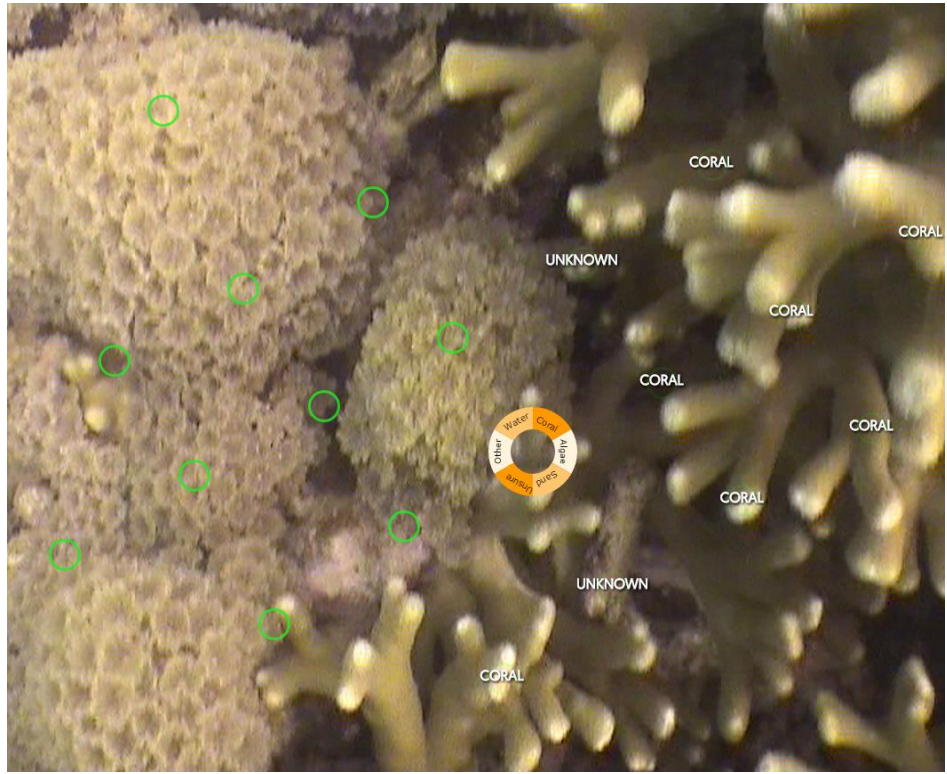

Figure 1. Partially classified citizen science image (0.12m$^2$), derived from Reef Check Australia videos. Green circles have yet to be classified, whereas white text are locations that have been classified. Each circle can be classified as water, (hard live) coral, algae, sand, unknown, or other.

2.5 Deriving weighted coral cover estimates

There were numerous discrepancies in the image-based coral-cover data to account for during the data integration process; including differences in sampling methodologies (Table 1), non-response rates (e.g. number of points classified as "unknown"), and disparities in classification accuracy between citizens, which adds to measurement uncertainty. Failing to account for these differences give data with high variances (e.g. high measurement uncertainty) too much weight in the model, leading to biased parameter estimates and confidence intervals (Gelman 2007). We chose to use weighted regression because it provides a natural way to account for these differences (Weisberg 2014). A detailed description of the deterministic weighting scheme applied to the citizen-science and the professional monitoring data are described in Sections 2.5.1 and 2.5.2, respectively. In Section 2.6, we describe the spatial data aggregation process used to generate the weighted coral-cover value for each reference raster cell and in Section 2.7, the weighted spatio-temporal Bayesian Beta regression model fit to the integrated dataset.

2.5.1 Weights for citizen-elicited coral cover measurements

The weight associated with the classification of image $j$, within source $s$, and classified by person $p$ is given by:

$$w_{jps} = w_{e_{js}} w_{n_{jps}} w_{a_p} w_{N_s}, \tag{1}$$

where $w_{e_{js}}$ is the physical extent of each image, $w_{n_{jps}}$ is the number of points used to elicit the coral cover in an image, $w_{a_p}$ is the overall accuracy of the person classifying the image, and $w_{N_s}$ is a factor used to up- or down-weight an image depending on source (Table 1). Each component contributing to $w_{jps}$ is described in more detail below.

2.5.1.1 Image extent

For each image and source, an extent weight, $w_{e_{js}} = A_j / max(A_j)$, was set based on the physical area of the image, $A_j$, captured by the camera. This is important because coral cover measurements are unitless proportions, but the extent over which they are calculated differs depending on the survey (Table 1). We standardized the weights based on the maximum image extent to ensure that $w_{e_{js}} \leq 1$.

2.5.1.2 Citizen's accuracy

We generated a classification accuracy weight, $w_{a_p}$, for each citizen based on their classification results for the 20 validation images annotated by the marine scientist. This was defined as the average accuracy across that person's classifications (Sammut and Webb 2010),

$$w_{a_p} = \frac{1}{\mathcal{J}_p} \sum_j \frac{\text{TP}_{jp} + \text{TN}_{jp}}{\text{TP}_{jp} + \text{TN}_{jp} + \text{FP}_{jp} + \text{FN}_{jp}}, j \in \mathcal{J}_p, \tag{2}$$

where $\text{TP}_{jp}$ is the number of true positives identified by citizen $p$ in image $j$, $\text{TN}_{jp}$ is the number of true negatives and similarly $\text{FP}_{jp}$ and $\text{FN}_{jp}$ are the false positives and negatives for each image and citizen, respectively. The set $\mathcal{J}_p$ represents the collection of images classified by person $p$.

2.5.1.3 Number of classification points and number of images

Citizens were asked to annotate 20 points on each of the 218 citizen science images, but individual responses labelled as "unknown" were removed from the analysis. Thus, the number of classifications per image for each person, $w_{n_{jps}}$, was sometimes less than 20. An additional factor, $w_{N_s}$, was used to account for the number of images used to generate a coral cover estimate (Table 1). However, $w_{N_s} = 1$ for all citizen contributed images in this study.

2.5.1.4 Weighted coral cover estimates: Citizen science data

Although classification points can be classified into numerous benthic categories, we are only concerned with whether a point was classified as hard coral or not. Thus, the coral cover estimate for a citizen's classification of an image from a specific source was the number of points, indexed $k$, that they labelled "coral" out of $w_{n_{jps}}$

$$y_{jps} = \frac{1}{w_{n_{jps}}} \sum_{k=1}^{w_{n_{jps}}} I(y_{jpsk} = \text{"coral"}), \tag{3}$$

with $I(\cdot)$ an indicator function and $y_{jpsk}$ the classification category label for point $k$ in image $j$ from source $s$ by citizen $p$.

Multiple people annotated the same image, but we ultimately wanted a single coral-cover estimate per image. Thus, the weighted coral-cover estimate, by image and source (i.e. Reef Check), was taken to be a weighted mean of the coral cover estimates derived from the participants,

$$\bar{y}_{js} = \frac{\sum_p w_{jps} y_{jps}}{w_{js}}, p \in P_{js}, \tag{5}$$

where $P_{js}$ represents the collection of people classifying the image and $w_{jps}$ is defined in (1). Note that this number is depicted as "variable" in Table 1 since it varies over images and sources. The total weight allocated to a measurement from a citizen-contributed image was

$$w_{js} = \sum_p w_{jps}, \ p \in P_{js}. \tag{4}$$

Thus, as more participants generate coral cover estimates for an image, the total weight for the citizen-science derived image, $w_{js}$, will increase, as desired.

### 2.5.2 Weights for professional monitoring data

The amount of information represented by the professional survey measurements differed depending on the survey design and classification procedures (Table 1). The image extent was accounted for in the same way it was for the citizen science images and we set $w_{a_p} = 1$ for all professional survey data, reflecting the increased ability of marine scientists to identify hard corals. The total number of points annotated on an image was included as $w_{n_j}$ (Table 1). The number of points required to obtain an automatic estimate of coral cover which is comparable to a manual classification by an expert is approximately 10 (Beijbom et al. 2015), and manual classification of additional points by a marine scientist within an image has not been shown to substantially improve coral cover estimates for an individual image. Therefore, the number of classification points allocated to the Catlin and UQ-RSRC coral cover estimates was $w_{n_j} = 10$, while the number of classification points for LTMP and MMP was set to 5. It was also necessary to include a weight to account for the number of images used to estimate coral cover, $w_{N_s}$, for the transect-level LTMP ($w_{N_s} = 40$) and MMP ($w_{N_s} = 32$), with $w_{N_s} = 1$ for all other sources (Table 1). This up-weighted these data to account for the fact that the measurements were aggregated over multiple images.

Thus, the weights associated with each estimate of coral cover from a professional survey, $w_{js}$, were derived as shown in Equation 1.

2.6 Spatial data aggregation

The weights described above provided a way to aggregate the various data sources into weighted measurements of coral cover. Next we spatially aggregated these observations to the reference-raster cell level for each source and year. It is common to spatially aggregate coral reef data prior to analyses (Gonzalez-Rivero et al. 2016; Sweatman et al. 2005) to separate true ecological patterns in space and time from noise (Habeeb et al. 2005). Aggregation also improves computational efficiency and ensures that the model is scalable as the number of images increases in the future. The spatial aggregation was accomplished by calculating the weighted-mean coral cover

$$\bar{y}_{its} = \frac{\sum_j w_{js} \bar{y}_{js}}{\sum_j w_{js}}, j \in \mathcal{J}_{its}, \qquad (6)$$

and the total weight

$$w_{its} = \sum_{j \in \mathcal{J}_{its}} w_{js}, \qquad (7)$$

for the collection of images, $\mathcal{J}_{its}$, that were contained within a cell, $i$, by source, $s$, and time, $t$. In total, there were 2056 cell-level estimates of coral cover. We wanted to model $\bar{y}_{its}$ using the Beta distribution (Section 2.7) and so we added the minimum weighted-mean coral cover greater than 0 (approximately 0.0005) to $\bar{y}_{its}$ to ensure that all values fell between 0 and 1.

To aid in interpretation and numerical stability, these weights were normalised prior to model fitting

$$w'_{its} = \frac{N \times w_{its}}{\sum_i w_{its}}, \qquad (8)$$

based on the sample size, $N$=2056, for each cell, time, and source.

2.7 Spatio-temporal modelling

Gaussian process models are often used for spatial statistical modelling and prediction, but these methods quickly become computationally infeasible for large datasets (Cressie and Wikle 2011). However, low rank approximations to Gaussian processes have been developed to address this issue, and freely available software has been developed to implement them. We chose to use a stochastic partial differential equation (SPDE) approach (Lindgren and Lindgren 2011) because the model's predictive accuracy is relatively high when data are spatially correlated and the prediction interval coverage reliably captures the true value when a nugget effect is included in the model. For an excellent overview of modelling methods for large spatial data and a comparison of their predictive performance, please see Heaton et al. (2018).

The proportion of coral cover, $\bar{y}_{its}$, within a cell, time, and source was modelled as a random draw from a Beta distribution with mean, $\mu_{its}$, and a common precision parameter, $\phi$, with a logistic link function and with the log likelihood, $L_{its}$, weighted by the normalised values generated in Eq. 8 and scaled to sum to 1. Hence the posterior distribution of the full parameter set, $\theta$, given the data, is given by

$$p(\theta|\bar{y}_{its}) \propto e^{\sum w'_{its} L_{its}}, \tag{9}$$

where

$$\bar{y}_{its} \sim \text{Beta}(\mu_{its}, \phi),$$

and

$$logit(\mu_{its}) = \mathbf{X}_{it}\boldsymbol{\beta} + u_i + v_t + \varepsilon_{its}$$

$$f_i(u_i) = GF(0, \Sigma).$$

The logit of the mean parameter was a function of a matrix of covariates, $\mathbf{X}_{it}$, and a vector of coefficients, $\boldsymbol{\beta}$, a temporal effect, $v_t$, fit using a first-order random walk, and a spatial random effect, $f_i(u_i)$, modelled using a Gaussian Markov random field with mean 0 and covariance, $\Sigma$, and fit using a SPDE. The triangulation nodes for the SPDE were based on the observed data locations and boundaries were constrained using the GBR Features shapefiles (GBRMPA 2014), with an extension radius of 0.5 dd. $\varepsilon_{its}$ was also included to represent uncorrelated errors. The spatio-temporal model was implemented using the r-INLA package (Rue et al. 2017) using the default priors. For $\phi$, this was a log-gamma prior with parameters 1.0 and 0.1, while the regression coefficients had a Normal prior with mean 0 and precision 0.001. A log-gamma prior was also used for the first-order random walk with parameters 1.0 and 0.00005. The SPDE is defined based on two parameters ($\kappa$ and $\tau$), which were modelled using a multivariate normal prior. The initial values for $\kappa$ and $\tau$ depend on the size of the mesh and in our case were set to -0.872 and -0.394, respectively. The covariates, $\mathbf{X}_{it}$, were always included in the model and no formal model selection was undertaken. All of the analyses were implemented in R statistical software (R Core Team 2017).

### 2.7.1 Prediction

The fitted model was used to generate coral cover predictions, $\hat{\mu}_{it}$, and corresponding standard deviations at each of the 85529 locations over time. The prediction locations were assigned a weight equal to the mean of the standardised weights generated in Eq. 8, which was 1.

### 2.7.2 Model Assessment

We fit the model to two separate datasets to test whether including additional data collected by multiple organisations (citizen science groups, universities, etc.), in addition to those collected by official monitoring programs increases the predictive ability of the models and thus, provides more accurate and precise information about coral cover across the whole-of-the GBR. First, we fit the model to all of the data available (i.e. LTMP, MMP, UQ-RSRC, Catlin, and Reef Check); we refer to this as the "All Data" model. Then, we fit the model to the LTMP and MMP data only; we refer to this model as the "LTMP/MMP Only" model.

We used a 10-fold cross-validation procedure to assess the predictive ability of the two models. Data from the UQ-RSRC, Catlin, and Reef Check surveys were randomly divided into 10 parts without respect for the spatio-temporal structure of the data. These validation data were iteratively removed from the training dataset before refitting both the All Data and LTMP/MMP Only models and making predictions at the validation sites. The models were compared based on the root mean square prediction error (RMSPE) of the observations versus cross-validation predictions, which is given by $\sqrt{\sum_{i=1}^{n} \frac{(\hat{y}_i - y_i)^2}{n}}$, where $\hat{y}_i$ is the prediction for the *i*th datum, $y_i$, after removing it from the observed dataset, and $n$ is the total number of observations used in the cross-validation procedure. The interpretation of the RMSPE is fairly straight forward. For example, if the RMSPE value for the best model is one quarter that of a competing model the gain in predictive ability is 75%. We also calculated the 95% prediction interval coverage, which is the percent of intervals containing the true value. For a single prediction, $q = \hat{\mu}_{it}$, the 95% prediction interval was calculated as $q \mp 1.96\sigma$.

### 2.7.3 Influence of citizen science data

There were relatively few benthic images available for this project from citizen scientists compared to professional data sources (Table 1). Yet, online platforms demonstrate the potential for citizen scientists to both contribute and process imagery. For example, the platform Zooniverse (Simpson et al. 2016) has tens-of-thousands of people who help process images and videos in scientific studies. It was important to understand how large volumes of citizen-contributed data would influence model results and so we undertook a simulation study to explore this issue. We treated the Catlin data as our "citizen-contributed" data because it contained a relatively large number of aggregated coral cover estimates collected at different locations in the GBR. We down-weighted the Catlin data to 1 on the standardised scale ($w'_{its}$) and refit the SPDE model to obtain predictions, with corresponding standard deviations. We repeated this three more times, but increased the normalised weights for the Catlin data from 1 to 1000, 10000, and 100000 at each iteration. The coral-cover measurement weights are a product of sub-weights and so this is an approximation of what would happen if the number of citizens classifying data was increased by 1000, 10000, and 100000. We refit the models using the adjusted weights and generated the RMSPE and 95% prediction coverage (proportion of observed values within the 95% prediction intervals).

3 **Results**

3.1 Classification of citizen science images

A total of 12 citizens annotated 218 Reef Check images. The average classification accuracy of citizens compared to the marine scientist for the 20 validation images was 79%, while the median accuracy value was even higher with the exception of one participant (Figure 2). This suggests that the users who annotated the images were more often than not correctly identifying image features that were, and were not, corals. However, there was also a considerable amount of variability in the classification accuracy of participants (Figure 2). This variability is likely attributable to image properties and the benthic composition (Appendix 3); the 20 Catlin images used in the training set were selected to capture a wide variety of reef characteristics, such as haziness, sand, and soft and hard corals and these characteristics increased or decreased the users' ability to accurately classify coral. We captured these differences in the individualised accuracy weights assigned to each person.

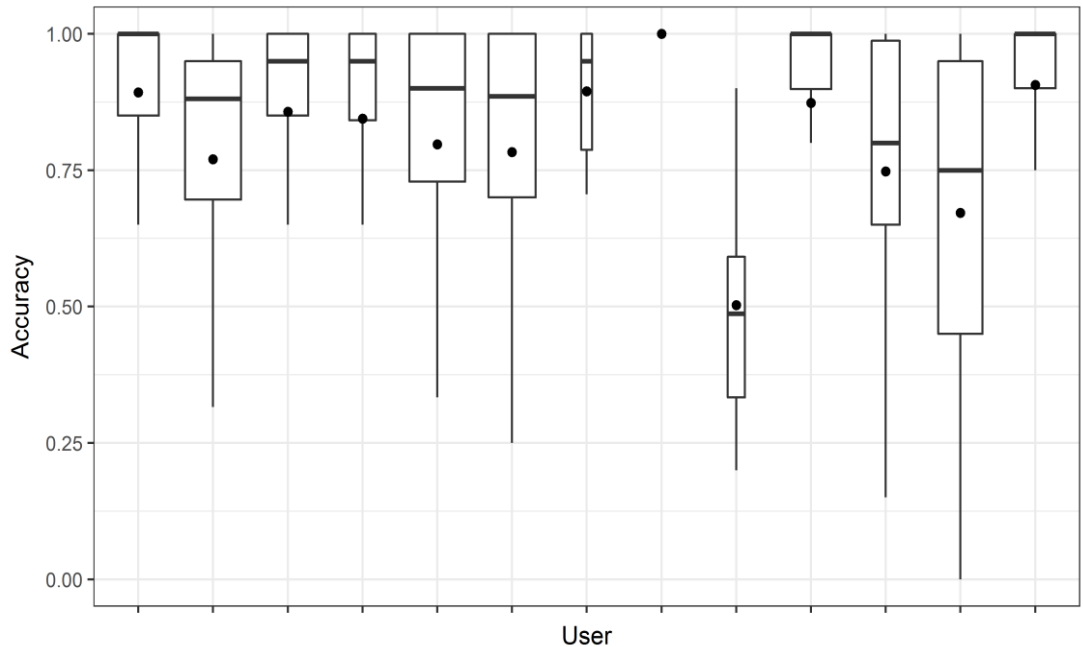

Figure 2. Boxplots show individual accuracy level of coral cover estimates obtained from 12 Citizens (User) compared to those of 1 marine scientist (whose accuracy is shown as a single dot at 1). Boxes delineate the 25[th] and 75[th] percentiles, black lines represent the 50[th] percentile, and black dots denote a user's mean accuracy across all images they annotated. The width of the boxplot is proportional to the square root of the number of classifications performed.

3.2   Exploratory analyses of hard coral cover data

After spatially aggregating the data, there were 2056 coral cover estimates available between 2002 and 2015. The number of cells that contained coral cover data, for each source and year is shown in Figure 3.

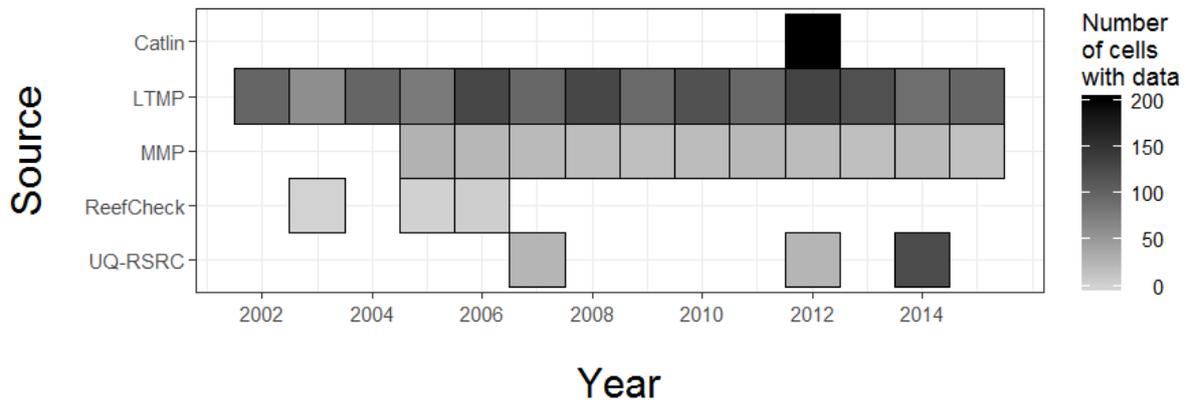

Figure 3. The number of reference raster cells containing at least one coral cover observation for each of the professional surveys and Reef Check citizen science data. Professional surveys included the Marine Monitoring Program (MMP), Long Term Monitoring Program (LTMP), the University of Queensland Remote Sensing Research Centre (UQ-RSRC) survey, and the XL Catlin Seaview Survey.

The normalised weights (Eq. 8) for the spatially aggregated data sources ranged between 0.0315 and 22.6717 for coral cover estimates derived from different sources (Figure 4). As expected, the aggregated coral-cover estimates from the Reef Check data were assigned the smallest weights (< 0.0947) because of the limited image extent (Table 1), the small number of citizens annotating the images (n=12), and the fact that citizens had accuracy weights less than 1 (Figure 2). Although the spatially aggregated coral-cover estimates for the LTMP and MMP were substantially up-weighted (Table 1) compared to other data sources, the coral cover estimates from the UQ-RSRC and Catlin surveys often had much larger weights (Figure 4) because large numbers of images were collected within the same cell (Appendix 1).

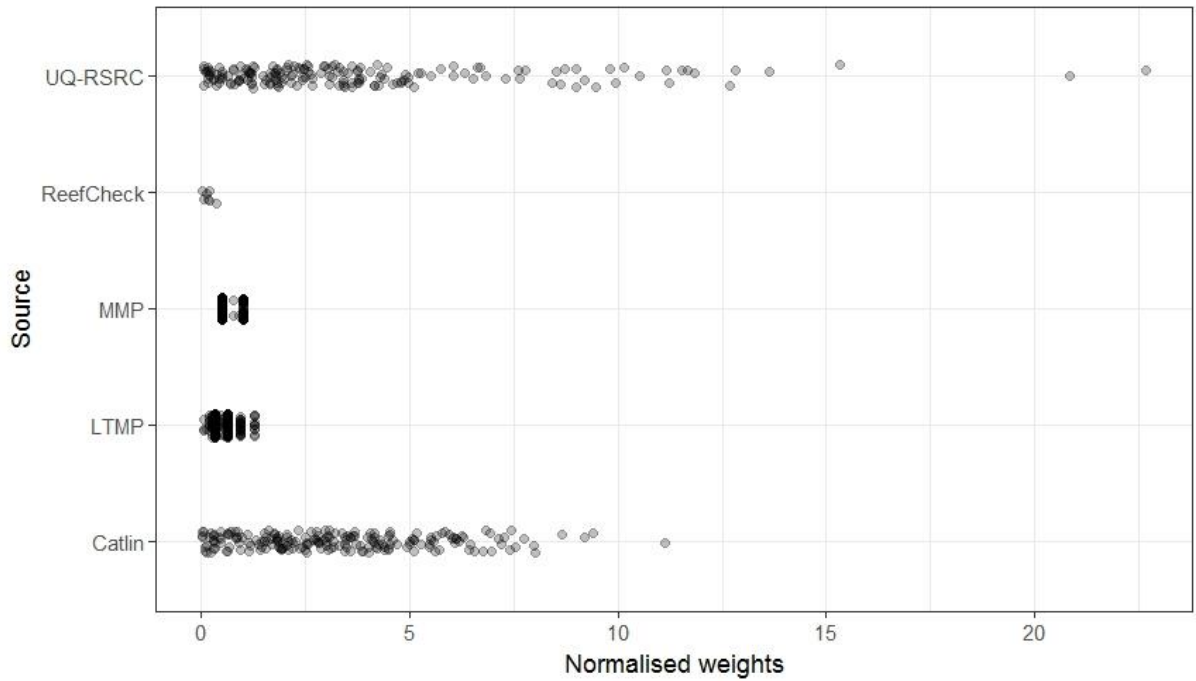

Figure 4. Normalised weights assigned to each coral cover estimate, by each of the professional surveys and Reef Check citizen science data. Professional surveys included the Marine Monitoring Program (MMP), Long Term Monitoring Program (LTMP), the University of Queensland Remote Sensing Research Centre (UQ-RSRC) survey, and the XL Catlin Seaview Survey. Differences in shading reflect overlaid points.

### 3.3 Model Results

As anticipated, the posterior estimate of the Beta precision term, $\phi$, was much larger for the All Data model than the LTMP/MMP Only model, reflecting the greater amount of data in the former. However, the fixed effects parameters in the two models were similar (Appendix 3, A3.4 and A3.5, respectively). Cyclone and bleaching damage were the only substantive covariates based on the 95% credible intervals and both negatively influenced coral cover, as expected. In addition, bleaching had the strongest negative influence in both models. Nevertheless, the relative reduction of the RMSPE between the two models indicates that the predictive ability of the All Data model was 42.76% better than the LTMP/MMP Only model and also better with respect to the 95% prediction coverage (Table 3). Both models tended to over-predict when observed coral-cover values were low and under-predict when they were high, but this was much more severe in the LTMP/MMP Only model (Figure 5; Appendix 3). As a result, there was a relatively strong positive relationship between the observed and cross-validation predictions in the All Data Model, but almost no relationship between the two in the LTMP/MMP Only model (Table 3; Figure 5).

Table 3. Results of the 10-fold cross-validation for the model fit to all the data (All Data) and the model fit to the LTMP and MMP data only, including the 95% prediction coverage, the root mean square prediction error (RMSPE), and the correlation (Corr) between the observed and predicted coral cover.

| Model | 95% Coverage | RMSPE | Corr |
| --- | --- | --- | --- |
| All Data | 86.47 | 0.0826 | 0.7431 |
| LTMP/MMP | 77.34 | 0.1443 | -0.0560 |

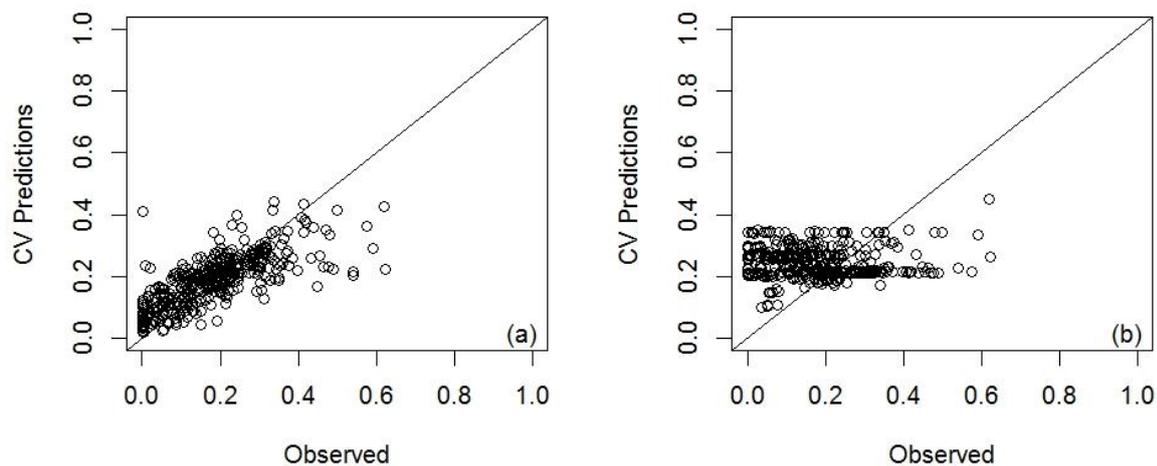

Figure 5. Plots of observed coral cover values versus the 10-fold cross validation predictions (CV Predictions) for the All Data model (a) and the LTMP/MMP Only model (b).

Animations of the model predictions and associated standard deviations for the All Data model are shown in Appendix 4, Figure A4.1 for years 2002-2015. The spatial variability in the predictions and corresponding standard deviations was greater in the All Data model compared to the LTMP/MMP Only model. It is not surprising that these patterns were found in areas where LTMP and MMP have not been collected because new data provide information about coral cover where none was previously present. However, this spatial variability was also found in areas where LTMP data have been collected, such as the area surrounding Heron Island (Figure 6a and Figure 7a). The LTMP has collected data in this area since 2002 (10 cells, n = 80), with the exception of 2005 and 2009. Additional data from the UQ-RSRC and Catlin surveys were also available in 2007, 2012, and 2014 for 127 cells. There was little spatial variability in the Heron Island predictions produced by the LTMP/MMP Only model (Figure 6b) compared to the All Data model (Figure 6c) and this was also true on reefs where no samples were collected. The standard deviation of the predictions increased in both models as the distance from observations increased, as is expected in a spatial statistical model (Figure 7). However, there was less spatial variability in the prediction standard deviations in the LTMP/MMP Only model (Figure 7b) relative those of the All Data model (Figure 7c). Interestingly, adding additional data did not decrease the uncertainty in the model predictions in all places (Figure 7c). In this case, the additional data included in the All Data model exhibited more spatial variability, which resulted in higher prediction standard deviations in nearby reefs where no data were available. Thus, these new estimates of uncertainty more accurately represent the uncertainty in the coral cover at those locations than those produced by the LTMP/MMP Only model.

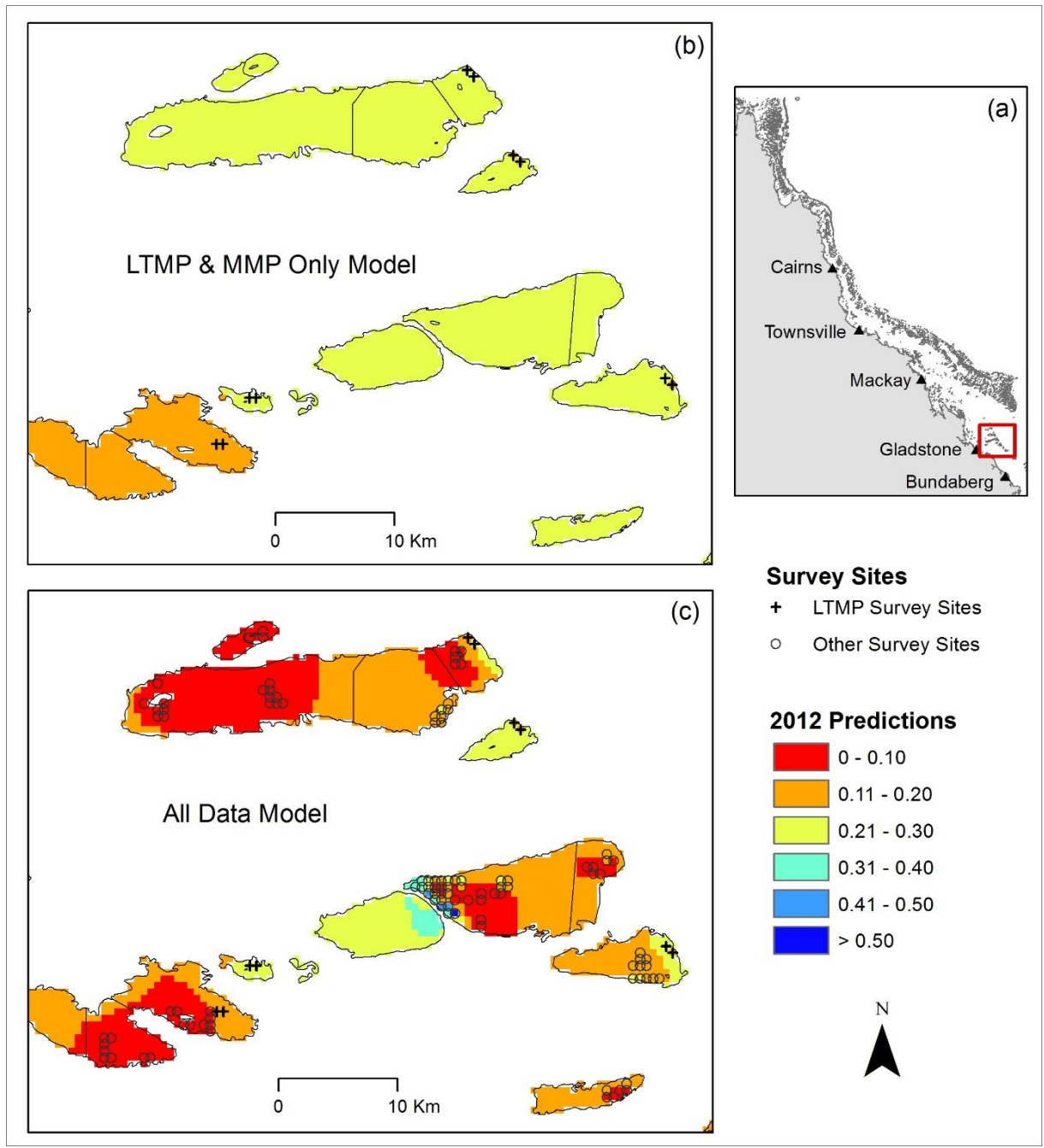

Figure 6. Predictions of coral cover for the Heron Island Region (a) generated by the All Data model (b) and the LTMP/MMP Only model (c) in 2012. Data at LTMP survey sites were collected in 2002-2004, 2006-2008 and 2010-2015. Data from the Other Survey Sites shown in (c) were collected by Catlin (2012) and the UQ-RSRC (2007, 2012, and 2014).

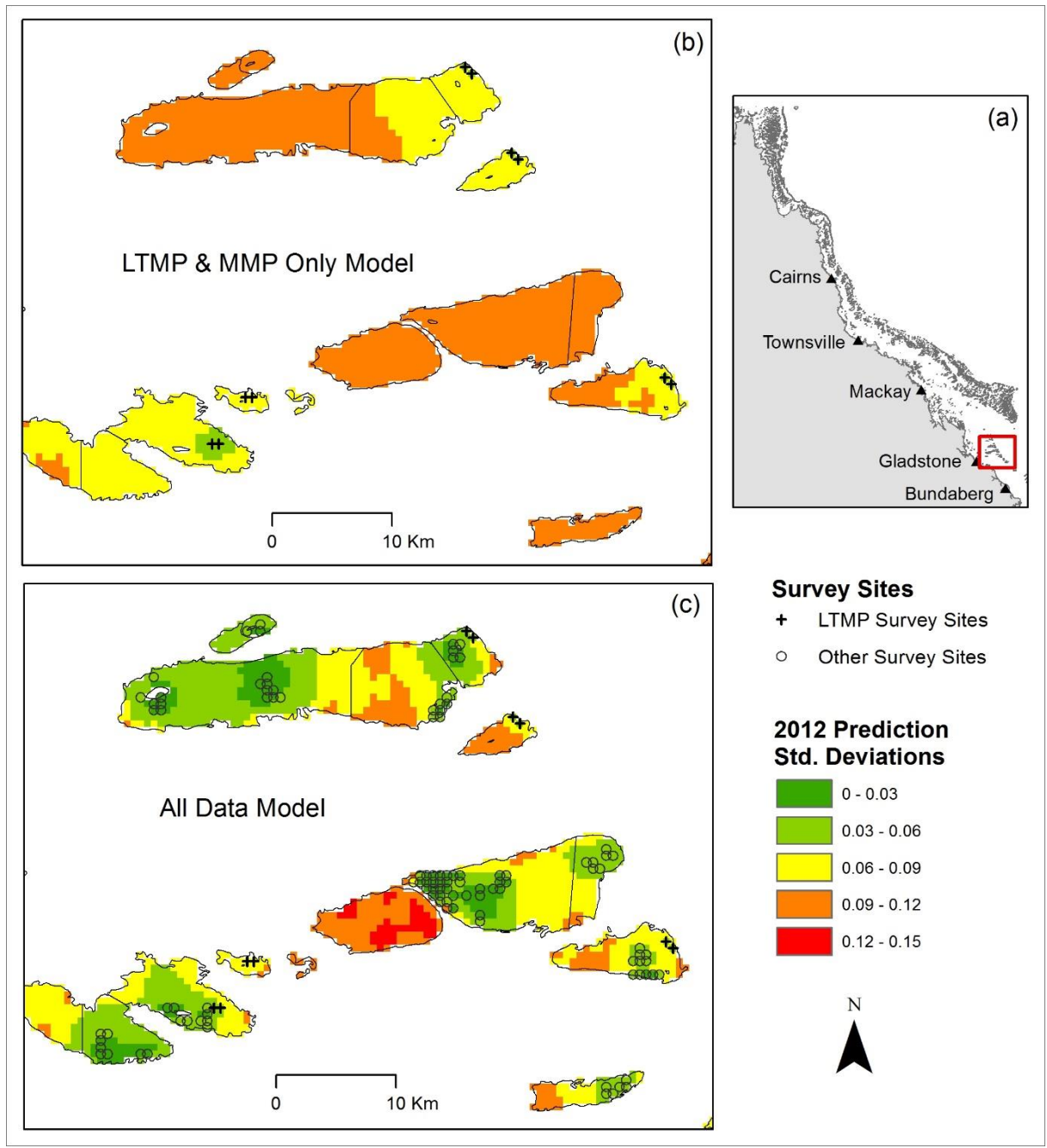

Figure 7. Prediction standard deviations for coral cover in the Heron Island Region (a) generated by the All Data model (b) and the LTMP/MMP Only model (c) in 2012. Data at LTMP survey sites were collected in 2002-2004, 2006-2008 and 2010-2015. Data from the Other Survey Sites shown in (c) were collected by Catlin (2012) and the UQ-RSRC (2007, 2012, and 2014).

As mentioned previously, the coral cover estimates from the pilot Reef Check study had little influence in the model because of the small number and extent of images used, as well as the small number of citizens classifying them. The simulation study allowed us to run a what-if scenario to assess the potential influence of citizen-contributed data as the number of classification points increases. The results showed that citizen-contributed data had no influence on model predictions and associated uncertainty estimates when there was a small number of images or small number of citizen classifications within a cell (Figure 8, Table 4). For example, when the weights were set to 1, there was no correlation between the observed and fitted citizen-science data values (Figure 8a), even though data from these exact locations and times were included in the spatio-temporal model (i.e. this was not a cross-validation procedure). When the weights were set to 1, the prediction intervals were exceptionally wide (Figure 8a), which means that the true coral cover value could fall anywhere within the prediction interval. This reflects the fact that we have little confidence in the quality of small amounts of citizen-contributed data and the subsequent model predictions based on those data. As the weights increased, the positive correlation between the observations and model predictions became progressively stronger and the prediction intervals narrower (Figures 8b,c,d), showing how the certainty in the model prediction grows as the number of citizen classifications within a cell increases. The 95% prediction intervals captured the true value at least 96% of the time regardless of the weights (Table 4), which means that the estimates of prediction uncertainty are reliable.

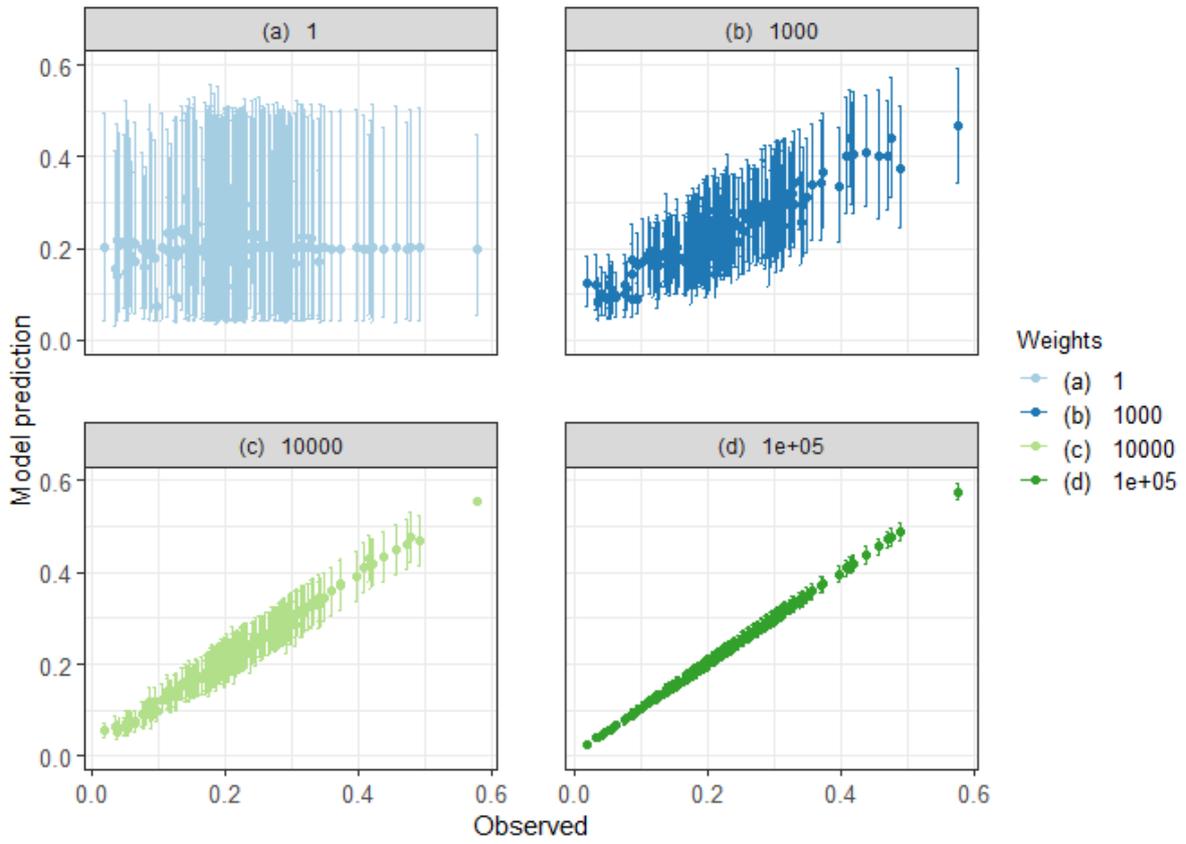

Figure 8. The predicted posterior mean coral cover values and 95% credible intervals for the simulated citizen-science data at the original scale (weight = 1), and then upweighted by 1000, 10000, and 100000 (1+e05).

Table 4. A summary of the predictive ability of the models fit to simulated citizen-science data at the original scale (weight = 1), and then upweighted by 1000, 10000, and 100000, including the 95% prediction coverage (95% Cov), as well as the root mean square error (RMSPE) and the correlation (Corr) between the observed and fitted values.

| Weights | 95% Cov | RMSPE | Corr |
|---|---|---|---|
| 1 | 0.9600 | 0.0968 | 0.1864 |
| 1,000 | 0.9650 | 0.0196 | 0.9885 |
| 10,000 | 0.9950 | 0.0029 | 0.9999 |
| 100,000 | 1.0000 | 0.0004 | 1.0000 |

## 4  Discussion

Data collection in the GBR is currently fragmented over dozens of organisations, with data collected for different purposes, using different methods, and in different habitats (e.g. reef slopes versus reef flats). As a result, data are rarely, if ever, analysed together (Hedge et al. 2013). And yet, there are numerous environmental applications where data from smart devices, *in situ* sensors, remote sensing, and professional monitoring efforts have been integrated within a modelling framework (Kadlec and Ames 2017; Granell et al. 2016; Hallgren et al. 2016; Wang et al. 2018). For example, Kadlec and Ames (2017) combined crowd-sourced snow reports, ski tracks from Strava and Garmin websites, data from meteorological stations, and remotely sensed imagery to significantly improve predictive maps of snow cover. However, most studies that integrate data sources either ignore the variable level of uncertainty in the measurements (Hallgren et al. 2016) or use a binary classification (i.e. certain versus uncertain) to identify and remove data sources deemed unreliable from the analysis (Kadlec and Ames 2017). Thus, the ability to integrate data collected at multiple scales, with variable levels of quality, remains one of the major challenges to implementing an 'Earth Observation Web' (Havlik et al. 2011), where a variety of uncertain data sources are combined within a modelling framework to provide spatially explicit and timely information to inform management (Granell et al. 2016).

We addressed this challenge using a mechanistically based weighting scheme within a spatio-temporal model to account for the differences in survey design and coral-cover estimation method from different image-based sources, as well as the uncertainty associated with different data sources. Integrating additional data in the All Data model did not negatively affect inferences about the relationship between coral cover and known pressures such as bleaching and cyclone damage (Appendix 3). In contrast, including additional data sources, in addition to the LTMP and MMP, significantly increased the predictive accuracy of the model (Table 3, Figure 5), even in areas where long-term datasets have been collected for more than 10 years (Figures 6 and 7). Thus, valuable information is lost when informative data from multiple sources are not integrated; information that could be used to support better data-enabled management decisions. This is not surprising given that the model contains a spatial statistical component (Eq. 9), which reduces parameter bias and improves predictive accuracy when data are spatially dependent (Hefley et al. 2017). Nevertheless, the improvement in predictive ability is significant given that only 384 additional measurements from 363 locations were included in the All Data model. As the volume and density of the data increases over time, the predictive accuracy of the spatial statistical model is also expected to increase; especially when new data come from areas that were not previously sampled.

Combining data from different sources, collected using different methods, with disparities in non-response rates (e.g. number of points not classified as 'unknown') and measurement uncertainty, and different sampling intensities is challenging, which is one reason these data are not usually analysed together. We used weighted regression because it is a natural way to account for these differences (Fuller 1987), but one drawback is that the weights are generally assumed to be "known" (Weisberg 2014), as was the case here. The mechanistic weights were specifically designed to account for measurable differences in the data related to survey and estimation method, classification non-response rates and accuracy, and the number of images used to estimate coral cover at an aggregated cell level (Table 1 and Section 2.7). However, it is impossible to verify that these are the optimal weights for each measurement. An alternative approach is to include covariates for the full weighting (e.g. image sources, extent, number of annotation points, and accuracy levels, and number of images within an aggregated cell), so that parameters can be estimated for each sub-weight. However, the model quickly becomes large and complicated, especially as the number of image sources and citizens classifying images increases. In addition, interactions must be included between sub-weight covariates, as well as other covariates in the model, and large numbers of interactions can lead to unstable estimates (Gelman 2007). Another way to address this issue in the future would be to extend the weighting scheme to allow for stochasticity in the weights, possibly by representing them as prior distributions within a Bayesian hierarchical model (Gelman 2007; Si et al. In Review). Depending on the parameterisation, this could also provide a way of incorporating measurement-specific precision, rather than using a pooled precision as we did here.

Reef managers and scientists often refrain from using citizen-science data due to a lack of confidence in the quality (GBRMPA 2013; Thompson et al. 2018); a common concern regardless of domain (Alabri and Hunter 2010; Bird et al. 2014; Burgess et al. 2016; Kosmala et al. 2016). It was therefore important to demonstrate how citizen-science data of potentially questionable quality can be integrated with professional monitoring data, without degrading model outputs. We used a combination of common approaches to address uncertainty in citizen-contributed data including professional verification (Gardiner et al. 2012; Kosmala et al. 2016); replication (Kosmala et al. 2016; Kyba et al. 2016; Swanson et al. 2016); spatio-temporal modelling to address biases in sampling effort, spatial clustering, and temporally dependent data (Bird et al. 2014; Isaak et al. 2014); and weighted regression to up- or down-weight data in the analyses (Anton et al. 2018; Kallimanis et al. 2017; Swanson et al. 2016; Mengersen et al. 2017). The average citizen's accuracy level was relatively high (79%), but there was substantial variability in their responses (Figure 2). Numerous studies have found that accuracy of citizen science data is related to the difficulty or complexity of the task (Aceves-Bueno et al. 2017; Kallimanis et al. 2017; Kosmala et al. 2016). In this model the inherent assumption is that each classification decision is equally difficult. However, a person's classification accuracy is likely to vary depending on image characteristics (e.g. haziness) or benthic composition (Appendix 3), and this could bias parameter estimates (Baker and Kim 2004). We did not attempt to account for variable levels of classification difficulty due to the small number of participants (12 citizen scientists) and test images (n = 20). However, elements of item response theory could be used to account for the difficulty in the classification decision (e.g. hard coral versus soft coral), in addition to a person's overall ability (Baker and Kim 2004; Fox 2010) as the amount of citizen-contributed data grows (e.g. Raykar et al. 2010).

There was little citizen-science data used in this study (~0.3%; Table 1) and as a result it had little weight in the model due to the limited pool of citizen scientists, the relatively small number of images, and the extremely small spatial extent of the Reef Check images used in this study (Table 1). Thus, a simulation study was used to understand how citizen's participation would influence model outputs as the amount of citizen-contributed data increased. The results showed that individual people, or small numbers of people, have almost no influence on the model results (Figure 8, Table 4). Instead, the weighting scheme ensures that large numbers of citizens (~1000) would need to classify each image before their aggregated data would have a strong influence on model predictions and associated estimates of uncertainty. Although these numbers seem large, they are not unreasonable given the participation rates in popular, online citizen-science studies (e.g. Simpson et al. 2016), where researchers rely on replication to help improve the quality of observations (e.g. Kyba et al. 2013; Swanson et al. 2016). For example, Kyba et al. (2013) analysed citizen's observations of artificial skyglow (e.g. light pollution) and found that aggregated data values from multiple observations provided a wealth of information, even though the uncertainty associated with individual observations was large. Thus, large quantities of moderate to low quality data have the potential to provide useful information (Reis et al. 2015).

Development of automated artificial intelligence (AI) algorithms for coral reef classification is rapidly increasing (Mahmood et al. 2017). However, the need for integration approaches will only increase as new algorithms are developed and implemented. Accuracy levels differ across AI algorithms and datasets due to differences in lighting conditions and water quality, morphological variability within species, similarities in the appearance of different soft and hard coral species, partial occlusion of objects due to the three-dimensional benthic structure, and difficulties defining the spatial borders between classes (Beijbom et al. 2015; Mahmood et al. 2017; Gómez-Ríos et al. 2019; González-Rivero et al. 2016). AI algorithms also require large training datasets of images classified by experts, whose classification accuracy varies, introducing errors into the algorithms (Beijbom et al. 2015; Gómez-Ríos et al. 2019). In addition, reductions in classification accuracy occur when algorithms are applied to datasets other than the one they were trained on (Gómez-Ríos et al. 2019). While there is no doubt that AI holds promise for broad-scale coral monitoring at fine spatio-temporal resolutions, there will continue to be a need to integrate data of variable quality across observers, regardless of whether those observers are AI algorithms, citizen scientists, and/or experts. This need will only grow as monitoring initiatives are implemented at broader scales, where data from different management regions or countries, and collected using different methods, must be integrated (e.g. Global Coral Reef Monitoring Network;

https://sustainabledevelopment.un.org/partnership/?p=14306).

**Management Implications**

A key goal of monitoring for conservation is to obtain information needed to guide effective decision-making (Nichols and Williams 2006). Within the GBR, RIMReP is tasked with coordinating and integrating monitoring data and model outcomes from numerous programs to 1) evaluate the management effectiveness of the Reef 2050 Plan (Commonwealth of Australia 2018), 2) report on status and trend, and 3) provide early warning on changes/impacts on the near- to mid-term horizon. Limited monitoring resources make it impossible for one organization to survey areas large enough, and with enough replication, to capture spatial and temporal heterogeneity within and between large regions (Nichols and Williams 2006; Yoccoz et al. 2001). Thus, the RIMReP must partner with organisations collecting data for short- to medium-term response monitoring, long term monitoring, and compliance monitoring, as well as citizen science programs in order to measure progress towards these outcomes (GBRMPA 2015). The data integration framework developed here provides a practical way to build on and align with existing programs, without duplicating effort or replacing them. In addition, the ability to integrate data from these different sources, while accounting for bias and uncertainty, promotes collaboration between government, researchers, industry and the community (GBRMPA 2015). Although we used coral cover as an example, this general approach to spatio-temporal data integration and modelling is equally viable for other variables collected in the marine environment, as well as in other ecosystems.

In a survey of GBR managers and other stakeholders, access to digital spatial data and predictive information via a web-based platform was overwhelmingly identified as their ideal information product (GBRMPA and Queensland Government 2018). The spatio-temporal data integration framework presented here can be used to provide the predictive information these stakeholders need, which is not possible using LTMP and MMP data alone. The models produce relatively fine-scale predictions across large areas within a probabilistic framework, with individual estimates of uncertainty, in areas where samples have not yet been collected (Cressie and Wikle 2011). This provides an integrated snapshot of coral cover condition across the GBR, based on the best available data at the time. However, the value of a predictive map depends on the prediction accuracy (i.e. how close are the predictions to the true value) and precision (i.e. how certain are we about the prediction). As mentioned previously, the prediction accuracy and precision is expected to increase as the spatial and temporal density of the data contributed by multiple organisations increases over time. This is a huge advantage in the GBR, where many organisations currently collect image-based coral cover data (Jonker et al. 2008; Ninio et al. 2003; González-Rivero et al. 2016; Roelfsema et al. 2018a; Roelfsema et al. 2018b), or could be collecting it in the future.

Predictive maps of coral cover generated from statistical models such as these add value to programs such as RIMReP because they provide insight into coral condition across the whole-of-the GBR. Predictive maps allow managers to explore spatial and temporal heterogeneity in coral cover to inform planning, assessment and decision making (Commonwealth of Australia 2015). For example, coral cover maps underpin other research studies related to fish abundance and community composition (Jones et al. 2004) or risk of disease outbreaks (Bruno et al. 2007). They can also be used as inputs for decision support tools; for example, connectivity modelling to identify optimal larval restoration sites (Hock et al. 2017). In addition, model predictions and associated uncertainty estimates can be summarized at different spatial scales (e.g. whole-of-the GBR, management regions, or reefs; Ver Hoef et al. 2008) and used to measure and report against targets in environmental report cards such as the 5-yearly Great Barrier Reef Outlook report (GBRMPA 2009), the annual Reef Water Quality Protection Plan Great Barrier Reef Report Card (Queensland Government 2017), and various regional report cards (e.g. Mackay-Whitsunday Healthy Rivers to Reef Partnership 2018).

Spatially explicit estimates of prediction uncertainty can also provide useful information for management. At the most basic level, differences in prediction uncertainty help people understand where they can be most and least confident in the predictions (Peterson and Urquhart 2006). This information could be used to prioritise management actions (Yoccoz et al. 2001), or identify areas where additional monitoring data are needed before management actions can be informed. Prediction uncertainty can also be used to inform the design of monitoring programs or guide future sampling efforts within and between organisations to ensure that monitoring resources are used in the most cost-effective way (Granell et al. 2016; Kang et al. 2016). In areas where the number of measurements is currently high, additional samples may not significantly impact the accuracy or the precision of the predictions. However, additional samples in areas where there are few or no observations can drastically change the model predictions and reduce estimates of uncertainty in those areas. These estimates of uncertainty can then be shared between organisations, so that sampling across the expanse of the GBR is better coordinated (Hedge et al. 2013), regardless of whether those collecting the data are professional monitoring teams or citizen scientists.

Traditional field surveys are expensive (Gardiner et al. 2012; Reis et al. 2015) and this is especially true in marine environments (Nygård et al. 2016; Lindenmayer and Likens 2010; Roelfsema and Phinn 2010). Thus, the ability to extract information from alternative data sources is financially attractive because it provides a cost-effective way to increase the spatio-temporal data coverage (Gardiner et al. 2012). However, these additional data must be viewed as complementary to government funded, professional monitoring data, rather than a replacement because there are trade-offs associated with data quality, the longevity of the survey, and financial costs (Gardiner et al. 2012; Reis et al. 2015). While government-funded professional monitoring programs such as the LTMP and MMP represent a large investment, they are currently the only way to ensure that high-quality long-term datasets exist (Figure 3) to assess long-term trends in coral cover and composition (Sweatman et al. 2005; Vercelloni et al. 2014). As we have shown here, these long-term datasets can be integrated with less certain, short- to medium-term spatially dispersed efforts by citizens. They can also be complemented by intensive monitoring efforts from research institutions (Roelfsema et al. 2018a; Roelfsema et al. 2018b), or groups funded by philanthropic organisations such as those from the XL Catlin Seaview Survey (González-Rivero et al. 2014), to assess spatial variability in coral cover. Government-funded monitoring programs are ideally suited to assess the impacts of planned management actions and RIMReP is an example where data from multiple sources is being integrated to achieve both short- and long-term sustainability targets (Addison et al. 2015). In the wake of broad-scale disturbances such as coral bleaching events (Hughes et al. 2018), however, coordinated efforts by multiple monitoring programs and institutions are required to provide rapid situational understanding. The modelling framework described here provides a way to integrate disparate data sources to maximize information value for management using all available data.

## 5 Conclusions

Monitoring efforts in the GBR are fragmented with dozens of organisations collecting data in different regions to meet a variety of monitoring objectives. We developed a weighted spatio-temporal Bayesian modelling framework that can be used to integrate and model multiple sources of coral cover data collected by both professional and non-professional organisations. There are numerous types of models that can be used to estimate coral cover, but there are a number of advantages to the model we describe here in terms of monitoring and management: 1) the mechanistic weighting scheme provides a way to integrate image-based data from multiple sources, while accounting for different levels of uncertainty; 2) the approach can be used with spatially and temporally dependent data collected for different purposes and using different survey designs; 3) the model produces spatially and temporally explicit predictions of coral cover, with estimates of uncertainty using a probabilistic framework; thus filling in gaps in space and time where no data exist; and 4) the accuracy of the predictions continue to improve as more data become available. The model results show clear advantages in terms of predictive performance and management; even in areas where professional datasets have been collected for more than 10 years. It also highlights the opportunities and potential for citizen science to contribute to data collection and processing. Predictive maps with estimates of uncertainty can be used to prioritize management actions, inform the design of monitoring programs and guide coordinated monitoring efforts within and across organisations, and provide scientifically rigorous information, summarised at multiple scales for environmental report cards. This spatio-temporal data integration approach can be applied to other variables and in different ecosystems, opening up opportunities to explore new ways to integrate data, enhance data processing power, and provide pathways for community engagement and shared stewardship.

## 6 Acknowledgements

This work has been supported by the Cooperative Research Centre for Spatial Information, whose activities are funded by the Business Cooperative Research Centres Programme. Would also like to thank the Queensland Department of Natural Resources Mines and Energy (DNRME), the Australian Research Council (ARC) Centre of Excellence in Mathematical and Statistical Frontiers (ACEMS), and the ARC Laureate program for the funding they provided for this research. Images and data were provided by Reef Check Australia; the University of Queensland (UQ) Global Change institute, Underwater Earth (previously The Ocean Agency), and the XL Catlin Seaview Survey; UQ Remote Sensing Research Centre;  and the Australian Institute of Marine Science. Special thanks to Reef Check Australia volunteers Paul Colquist, Douglas Stetner, Cheryl Tan, Hannalena Vaisanen, and Nathan Caromel, who classified hard coral for this study. Thanks to Sam Matthews for sharing spatial data representing coral bleaching and CoTS density at the GBR scale and to Andrew Zammit-Mangion for spatio-temporal modelling advice. Finally, we thank three anonymous Reviewers for their constructive comments, which helped improve the manuscript.